\documentclass[aps,prl,twocolumn,superscriptaddress]{revtex4-1}
\usepackage{amsmath}
\usepackage{graphicx}
\usepackage{color}
\usepackage{gensymb}
\usepackage{braket}
\usepackage{hyperref}
\usepackage{upgreek}

\usepackage{soul}

\begin{document}

\title{Acoustically driven magnon-phonon coupling in a layered antiferromagnet}
	
\author{Thomas P. Lyons}
\affiliation{Center for Emergent Matter Science, RIKEN, Wako-shi, Saitama 351-0198, Japan}
\author{Jorge Puebla}
\email{jorgeluis.pueblanunez@riken.jp}
\affiliation{Center for Emergent Matter Science, RIKEN, Wako-shi, Saitama 351-0198, Japan}
\author{Kei Yamamoto}
\affiliation{Advanced Science Research Center, Japan Atomic Energy Agency, Tokai, Ibaraki 319-1195, Japan}
\affiliation{Center for Emergent Matter Science, RIKEN, Wako-shi, Saitama 351-0198, Japan}
\author{Russell S. Deacon}
\affiliation{Advanced Device Laboratory, RIKEN, Wako-shi, Saitama 351-0198, Japan}
\affiliation{Center for Emergent Matter Science, RIKEN, Wako-shi, Saitama 351-0198, Japan}
\author{Yunyoung Hwang}
\affiliation{Center for Emergent Matter Science, RIKEN, Wako-shi, Saitama 351-0198, Japan}
\affiliation{Institute for Solid State Physics, University of Tokyo, Kashiwa, Chiba 277-8581, Japan}
\author{Koji Ishibashi}
\affiliation{Advanced Device Laboratory, RIKEN, Wako-shi, Saitama 351-0198, Japan}
\affiliation{Center for Emergent Matter Science, RIKEN, Wako-shi, Saitama 351-0198, Japan}
\author{Sadamichi Maekawa}
\affiliation{Center for Emergent Matter Science, RIKEN, Wako-shi, Saitama 351-0198, Japan}
\affiliation{Advanced Science Research Center, Japan Atomic Energy Agency, Tokai, Ibaraki 319-1195, Japan}
\affiliation{Kavli Institute for Theoretical Sciences, University of Chinese Academy of Sciences, Beijing 100049, People’s Republic of China}
\author{Yoshichika Otani}
\affiliation{Center for Emergent Matter Science, RIKEN, Wako-shi, Saitama 351-0198, Japan}
\affiliation{Institute for Solid State Physics, University of Tokyo, Kashiwa, Chiba 277-8581, Japan}
	
\date{\today}

\begin{abstract}
    Harnessing the causal relationships between mechanical and magnetic properties of van der Waals materials presents a wealth of untapped opportunity for scientific and technological advancement, from precision sensing to novel memories. This can, however, only be exploited if the means exist to efficiently interface with the magnetoelastic interaction. Here, we demonstrate acoustically-driven spin-wave resonance in a crystalline antiferromagnet, chromium trichloride, via surface acoustic wave irradiation. The resulting magnon-phonon coupling is found to depend strongly on sample temperature and external magnetic field orientation, and displays a high sensitivity to extremely weak magnetic anisotropy fields in the few~mT range. Our work demonstrates a natural pairing between power-efficient strain-wave technology and the excellent mechanical properties of van der Waals materials, representing a foothold towards widespread future adoption of dynamic magneto-acoustics.
\end{abstract}
	
	
\maketitle


\begin{figure*}
    \center
    \includegraphics[scale=1]{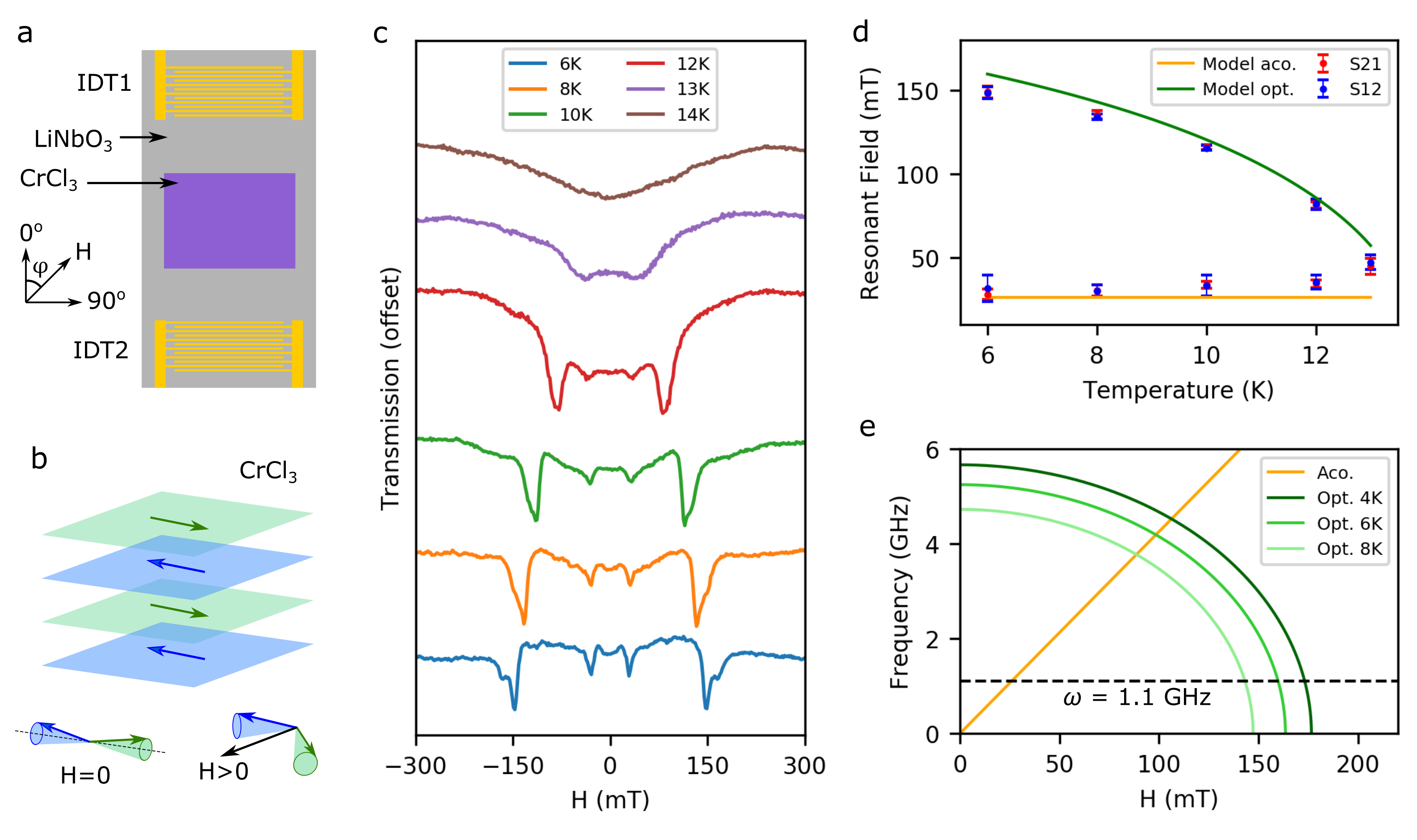}
    \caption{\label{fig:figure1} \textbf{Magnon-phonon coupling in layered CrCl$_3$.} (a) Schematic of the devices used in this work. See text for description. (b) CrCl$_3$ consists of stacked ferromagnetic layers of alternating in-plane magnetization, represented by two spin sublattices (green and blue arrows). In the absence of an external magnetic field, the sublattice magnetizations point away from each other, while an applied field causes them to cant. In-phase and out-of-phase precession of the sublattice magnetizations are associated with acoustic and optical magnon modes, respectively. (c) SAW transmission signal through CrCl$_3$ in Sample 1 as a function of applied magnetic field strength at an angle $\phi = 45\degree$, at various sample temperatures. (d) Extracted resonance field strengths for the acoustic and optical magnon modes at various Sample 1 temperatures. Overlaid curves are calculated from the model described in the text. (e) Calculated frequency dependence of the acoustic and optical magnon modes as a function of applied magnetic field, at $T = 4, 6$ and 8~K.}
\end{figure*} 

\begin{figure*}
    \center
    \includegraphics[scale=1]{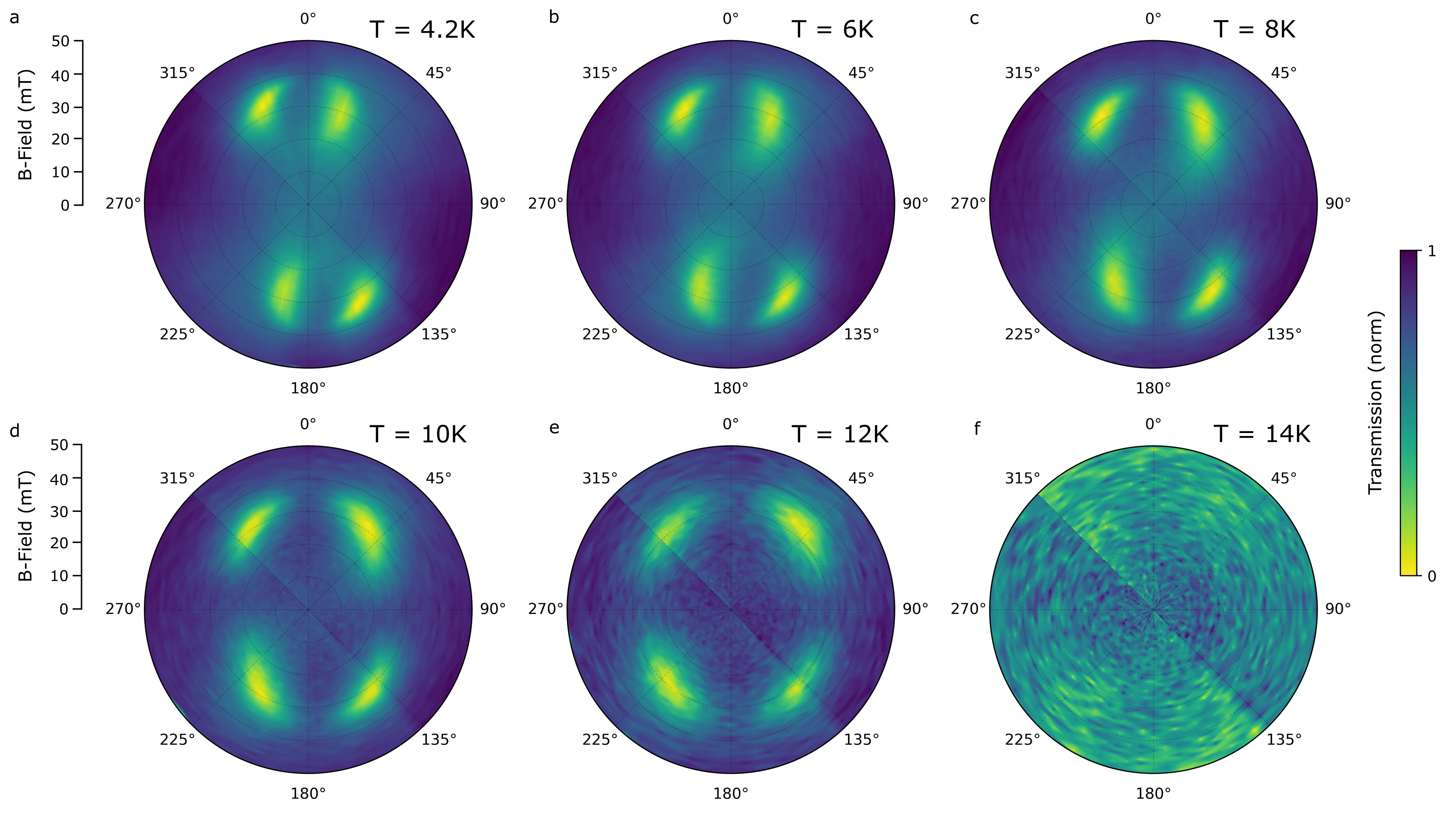}
    \caption{\label{fig:figure2} \textbf{Acoustic magnon mode dependence on external field angle and temperature} (a-f) Polar plots of SAW absorption by the acoustic magnon mode in Sample 2, at various sample temperatures, as a function of applied external magnetic field orientation in the sample plane. Asymmetry at lower temperatures arises due to very weak uniaxial anisotropy $\sim 2$~mT. Upon heating, the expected symmetric response of the magnetoelastic interaction is recovered. Absorption disappears at $T = 14$~K, close to the N\'eel temperature. }
\end{figure*} 

\begin{figure}
    \center
    \includegraphics[scale=1]{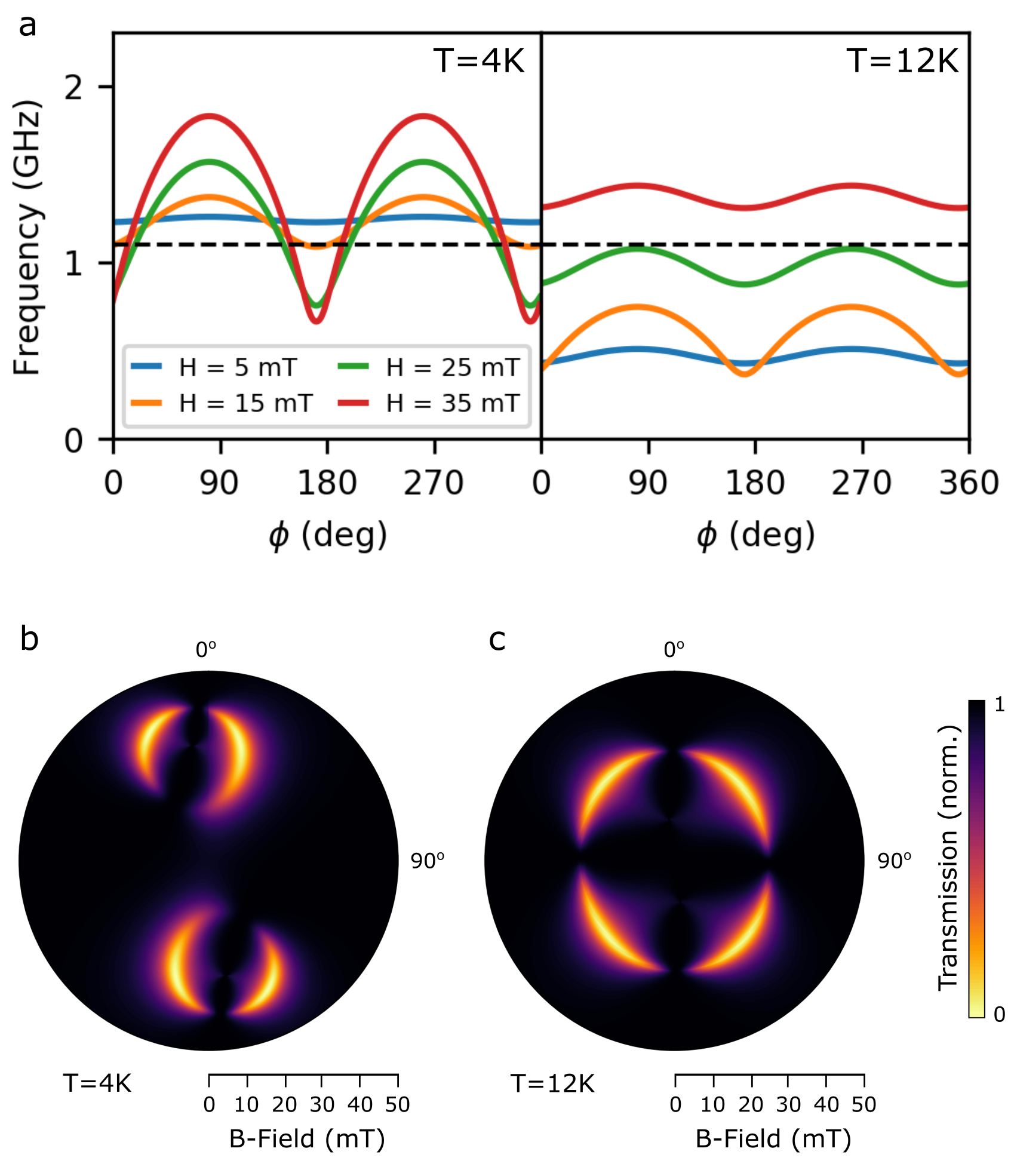}
    \caption{\label{fig:figure3} \textbf{Theoretical model for acoustic mode} (a) Calculated acoustic mode frequency dependence on external magnetic field orientation $\phi$. (b, c) Simulated polar plots of SAW absorption by the acoustic magnon as a function of external magnetic field orientation, using parameters for Sample 2. The striking difference in response is largely attributed to a change in anisotropy of only $\sim 1$ mT.}
\end{figure}


From uncertain beginnings, the technological advantages of antiferromagnets over ferromagnets are now well known, including fast operation, immunity against device crosstalk and stray fields, and amenability to low power control via spin currents or proximitized materials~\cite{Jungwirth2016,Fukami2020}. However, these very same advantageous properties can be a double-edged sword, being partly responsible for a general lack of understanding of antiferromagnets as compared to ferromagnets. The high spin-wave frequencies can be prohibitive for probes based on microwave electronics, while the insensitivity to measurement techniques such as SQUID magnetometry or the magento-optical Kerr effect limit the effectiveness of these popular conventional magnetic probes. A less well-known probe, which has proven itself useful in the study of ferromagnets, relies not on optical or direct magnetic sensing but instead employs the magnetoelastic interaction between spin-waves and acoustic waves~\cite{Puebla2020,Yamamoto2022}. When in contact with a piezoelectric material, the magnetic film can be irradiated with surface acoustic waves (SAWs). Beyond the magnetic film, the transmitted SAWs can be measured, providing information on the magnet's response to external stimuli~\cite{Puebla2020,Xu2020}. Aside from the energy efficient generation, inherently low attenuation, suitability for miniaturization and long distance propagation of SAWs~\cite{Dreher2012,Nie2022,Puebla2020}, a particular advantage of this technique is that it does not discriminate between ferromagnetic and antiferromagnetic order, and indeed may even be stronger for the latter~\cite{Yamada1966}.

SAW technology is relatively mature, having found multiple applications in the microelectronics industry, yet continues to play a key role at the forefront of fundamental research, with recent notable advances including SAW-driven transport of single electrons in gallium arsenide~\cite{Wang2022}, semiconductor interlayer excitons in van der Waals heterobilayers~\cite{Peng2022}, and manipulation of the charge density wave in layered superconductors~\cite{Yokoi2020}, amongst other advances~\cite{Nie2022}. Utilizing SAWs as a probe of ferromagnetism has proven highly effective, for instance in understanding the fundamentals of magnetoelasticity and magnetostriction, or more recently in revealing the various mechanisms of SAW nonreciprocity~\cite{Xu2020,Puebla2020,Sasaki2017,Hernandez2020,Verba2019,Kuss2020,Kuss2021,Piyush}. Such works have laid the foundations for the active field of SAW-spintronics, in which dynamically applied strain can modulate magnetic properties~\cite{Dreher2012,Puebla2022}. This technique is mature for ferromagnets, and has recently been proven effective for multiferroics~\cite{Sasaki2019} and synthetic antiferromagnets~\cite{Matsumoto2022,Verba2019}, but a demonstration of SAW-driven magnon-phonon coupling in a crystalline antiferromagnet remains elusive.

Here, we utilize SAWs to drive spin-wave resonance in a layered crystalline antiferromagnet, chromium trichloride~(CrCl$_3$), a material characterized by layers of alternating magnetization weakly bound by van der Waals attraction~\cite{McGuire2017,Narath1965}. The antiferromagnetic order occurs only between adjacent monolayers rather than within them, giving rise to relatively weak interlayer exchange and associated lower frequency range of spin excitations in CrCl$_3$ as compared to conventional antiferromagnets~\cite{Macneill2019,McGuire2017}. The combination of easy flake transfer onto arbitrary substrates, with sub-10 GHz spin excitations, is advantageous for integration of CrCl$_3$ into SAW devices, where antiferromagnetic magnetoelasticity can be probed directly. After first demonstrating acoustic antiferromagnetic resonance, we proceed to study the influence of temperature and angle of applied external magnetic field on the magnon-phonon coupling. The sets of experimental data are analyzed by extending the established theoretical model for SAW-spin wave coupling in ferromagnetic films~\cite{Yamamoto2022,Xu2020}. Combined with a mean-field calculation of the temperature dependence, our model reproduces the observed features well, confirming the amenability of SAWs as a powerful probe to elucidate the dynamics of van~der~Waals magnets, especially given their excellent plasticity~\cite{Cantos2021}. Considering also that acoustic magnetic resonance generates spin currents, which have been shown to travel over long distances in antiferromagnets, our results offer an alternative route towards novel spintronic devices with layered crystals~\cite{awschalom2021,Lebrun2018,Burch2018,Ghiasi2021}.

Two devices are studied in this work. They each consist of lithium niobate (LiNbO$_3$) substrates with aluminium interdigital transducers (IDTs) either side of a CrCl$_3$ flake (Fig.~\ref{fig:figure1}a). Each IDT, 1 or 2, can generate SAWs at 1.1~GHz and wavelength 3.2~$\upmu$m, which subsequently propagate along the surface of the LiNbO$_3$, interact with the CrCl$_3$ flake, and then reach the other IDT where they are detected. By measuring SAW transmission in this way, any absorption of acoustic energy by the antiferromagnet can be detected (see methods). Sample~1 is quasi-bulk, at $\sim 4 \; \upmu$m thick, while Sample~2 is much thinner at $\sim 120$~nm (see Supplementary Information (SI)).

Below the N\'eel temperature of $\sim14$~K, layered CrCl$_3$ is composed of stacked ferromagnetic layers ordered antiferromagnetically~\cite{McGuire2017,Narath1965}. Alternate layers belong to one of two spin sublattices oriented collinearly in the layer plane, owing to easy plane anisotropy of strength $\sim250$~mT (Fig.~\ref{fig:figure1}b)~\cite{McGuire2017}. Two magnon modes arise from in-phase or out-of-phase precession of the two sublattice macrospins, described as acoustic and optical modes, respectively~\cite{Macneill2019}. In our experiments we apply an external magnetic field perpendicular to the crystal c-axis, inducing the two spin sublattice magentizations to cant towards the applied field direction (Fig.~\ref{fig:figure1}b). Such noncollinear canting modifies their precession frequency, thereby bringing the magnon modes into resonance with the acoustic wave.

We first apply an external magnetic field at an angle $\phi = 45\degree$ to the SAW propagation direction in Sample~1, and measure the amplitude of the SAW transmission. The result is shown in Fig.~\ref{fig:figure1}c, where clear transmission dips can be seen arising from absorption of SAWs by magnons. At $T=6$~K, absorption is observed at approximately 30 and 150~mT, attributed to the acoustic and optical modes, respectively. Examples of other external field orientations can be seen in the SI. Upon heating the sample, the optical mode absorption shifts to lower resonance field strengths while the acoustic mode stays largely insensitive to temperature (Fig.~\ref{fig:figure1}d). At $T=13$~K, the two modes are no longer resolved, and at $T=14$~K, close to the N\'eel temperature~\cite{McGuire2017}, they have disappeared. 


The observed temperature dependence of the resonance field can be modelled by combining a simple mean-field theory with the known formulae for spin wave resonance in easy-plane antiferromagnets~\cite{Macneill2019}
\begin{equation}
    H_{\rm res} = \begin{cases}
    \sqrt{2H_E /(2H_E+M_s)} \omega /\gamma & {\rm acoustic} \\
    \sqrt{4H_E^2 - 2H_E \omega ^2 / (M_s\gamma ^2 )} & {\rm optical} 
    \end{cases}
\end{equation}

\noindent Here $H_E $ is the interlayer exchange field, $M_s$ is the saturation magnetization, $\omega $ is the SAW frequency, and $\gamma /2\pi = 28$~GHz/T is the gyromagnetic ratio respectively. We solve the molecular field equation self-consistently in the macrospin limit $S\rightarrow \infty $ to obtain $M_s ( T)$. This approximation also implies $H_E (T) \propto M_s (T)$, which predicts the optical mode resonance field tends towards zero as the N\'eel temperature is approached while the acoustic mode remains unchanged. The calculated temperature dependence is plotted in Fig.~\ref{fig:figure1}d and agrees well with the experimental data. The small increase of the observed acoustic mode resonance field towards higher temperature~\cite{Zeisner2020} points to breakdown of the mean-field approximation near the phase transition. The same model can be used to calculate the effective magnon frequency evolution as a function of applied magnetic field strength, as shown in Fig.~\ref{fig:figure1}e.

We now consider the coupling between SAWs and the acoustic magnon mode in greater detail. Figure~\ref{fig:figure2} shows absorption by the acoustic mode as a function of external magnetic field orientation in the plane of Sample~2, where the vertical axis (0$\degree$ - 180$\degree$ line) is the SAW propagation axis. At $T=4.2$~K, we observe four lobes of strong absorption, seen only when the external magnetic field is applied at angles smaller than 45$\degree$ to the SAW propagation axis. As the temperature is increased to $T=12$~K, they migrate to new positions which are more rotationally symmetric. By $T=14$~K, close to the N\'eel temperature, the absorption has disappeared, in agreement with Sample~1.

To fully understand the results in Fig.~\ref{fig:figure2}, we must consider the interplay between antiferromagnetic resonance and magnon-SAW coupling. Each has its own dependence on external magnetic field orientation, with the latter defining the window through which we can observe the former. Firstly we focus on the magnetic response of CrCl$_3$ itself. Close inspection of Fig.~\ref{fig:figure2}a reveals that not only the magnitude of absorption but also the resonance field depends strongly on the magnetic field angle $\phi $ at $T=4.2$~K, indicating the presence of magnetic uniaxial anisotropy. To reproduce this observation, we calculate the acoustic mode resonance frequency as a function of $\phi $ computed for a model that includes an in-plane uniaxial anisotropy field $\mu _0 H_u \approx 2.1$~mT, oriented approximately along the line 171$\degree$ - 351$\degree$. Although this anisotropy is itself very weak, it induces a sizable zero-field magnon frequency gap of $\gamma \mu _0\sqrt{2H_u (2H_E +M_s + H_u )} \sim 1.2$~GHz, above the SAW frequency of 1.1~GHz. As can be seen at $T=4$~K in Fig.~\ref{fig:figure3}a, for $30\degree \lesssim \phi \lesssim 130\degree$ and $210\degree\lesssim \phi \lesssim 310\degree$, the frequency monotonically increases as $H$ increases so that the acoustic magnon never becomes resonant with the SAWs. Only in the remaining angular ranges are acoustic spin-wave resonances observable, which correspond to the lobes in Fig.~\ref{fig:figure2}a.

According to the well-known formula $H_u (T) \propto M_s (T)^2$~\cite{Callen1966}, the uniaxial anisotropy tends to zero as $T$ increases towards the N\'eel point. We find it reduces to $\approx 0.6$~mT at $T=12$~K, lowering the zero-field magnon frequency below the SAW frequency, and thereby allowing acoustic magnon resonance at 1.1~GHz for all angles at around $25-30$~mT (Fig.~\ref{fig:figure3}a).  While uniaxial anisotropy of $\sim 1$~mT has been observed before in CrCl$_3$~\cite{Kuhlow1982}, the origin remains ambiguous. Here, we tentatively ascribe it to negative thermal expansion in CrCl$_3$, in which the $a$-axis lattice constant gradually increases upon cooling the crystal below $T = 50$~K, owing to magnon induced expansion of the lattice~\cite{Schneeloch2022,Liu2022}. Our results hint at the applicability of SAWs to further investigate this poorly understood effect, or moreover exploit it for highly sensitive static strain or force sensing applications.

To complete the picture, we now consider the magnon-SAW coupling dependence on external field orientation, which has proven the key to accessing various parameters in ferromagnetic materials~\cite{Xu2020}. Given that, unlike ferromagnets, the antiferromagnetic sublattice magnetizations do not simply align with the external field, we model the magnetoelastic coupling in CrCl$_3$ by a free energy density $F_{\rm me} = b\epsilon _{ab}( n_a^A n_b^A + n_a^B n_b^B ) +2c\epsilon _{ab}n_a^A n_b^B  $. Here $\epsilon _{ab}$ is the strain tensor, $n_a^A , n_a^B$ are components of the normalized sublattice magnetization vectors, and Einstein's summation convention is assumed. $b$ is an intrasublattice magnetoelastic coefficient, a direct generalization of the ferromagnetic magnetoelasticity. $c$ is an intersublattice coefficient, unique to antiferromagnets, which was studied in literature~\cite{Borovik-Romanov1965}. Let $\phi _A , \phi _B$ be the angles between the SAW propagation direction and the respective sublattice magnetizations. The corresponding magnon-SAW couplings $g_A ,g_B$ exhibit the following angle dependence (see SI):

\begin{align}
     g_A \propto &\ b \sin \phi _A \cos \phi _A + c \sin \phi _A \cos \phi _B , \label{eqn:magnonSAW1} \\
    g_B \propto &\ b \sin \phi _B \cos \phi _B + c \sin \phi _B \cos \phi _A . \label{eqn:magnonSAW2}
\end{align}

\noindent The acoustic and optical modes see $g_A \pm g_B $ respectively, reflecting the phase relations between the two sublattices. For acoustic mode resonance, $H$ is small so that $\phi _B \approx \phi _A + \pi \approx \phi \pm \pi /2$, yielding $g_A +g_B  \propto \sin 2\phi $. This acoustic magnon-SAW coupling filters the nominally observable resonance frequencies shown in Fig.~\ref{fig:figure3}a to give the cumulative responses shown in Fig.~\ref{fig:figure3}b, c, in which vanishing absorption can be seen at $\phi =0\degree, 90\degree, 180\degree ,270\degree$. The agreement with Fig.~\ref{fig:figure2}a, e is satisfactory.

\begin{figure}
    \center
    \includegraphics[scale=1]{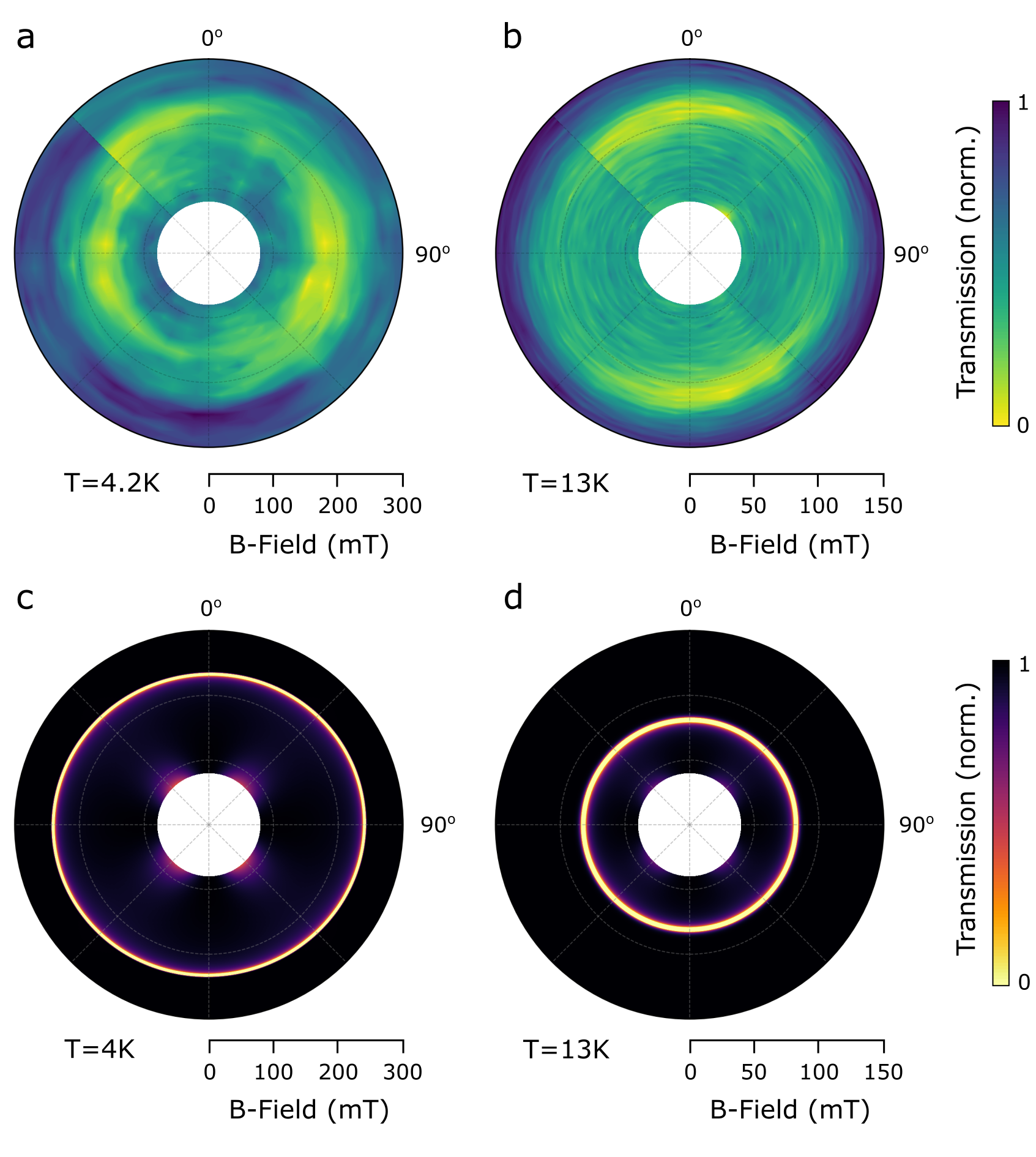}
    \caption{\label{fig:figure4} \textbf{Optical magnon mode dependence on external field angle and temperature} (a,b) Experimental and (c,d) simulated polar plots of SAW absorption by the optical mode in Sample 2 as a function of external magnetic field orientation at $T = 4$~K (experimental base temperature $T = 4.2$~K) and 13~K.}
\end{figure}

Next, we consider optical magnon-phonon coupling. Figures.~\ref{fig:figure4}a, b show the optical mode absorption in Sample~2, seen to some extent at every angle of applied field. This isotropic behaviour, in stark contrast to that displayed by the acoustic mode, arises because the two canted spin sublattices adopt an almost parallel configuration at the relatively high field strength needed to reach resonance, i.e. $\phi _A \approx \phi + \delta , \phi _B \approx \phi -\delta , |\delta |\ll \pi$. Equations~(\ref{eqn:magnonSAW1}) and (\ref{eqn:magnonSAW2}) therefore yield $g_A -g_B \propto \left( b \cos 2\phi + c \right) \sin 2\delta $. We note that the intrasublattice coupling $b$ alone gives a vanishing absorption at $\phi =45\degree$, inconsistent with both Sample~1 (Fig.~\ref{fig:figure1}c) and Sample~2 (Fig.~\ref{fig:figure4}a, b). Hence we take $b=0, c \sim 10^6$ J/m$^3$ with the aforementioned temperature dependent $H_E ,M_s ,H_u$ to generate Figs.~\ref{fig:figure4}c, d, which show the simulated optical mode absorption at $T=4$~K and 13~K, respectively. The agreement with experiment is satisfactory at $T=4.2$~K, and reasonable at $T=13$~K, given the simplifications to the model (such as an absence of broadening/disorder) and the expected breakdown of the mean-field approximation close to the phase transition.

In conclusion, we demonstrate GHz-range SAW-driven magnon-phonon coupling in a crystalline antiferromagnet. This demonstration paves the way towards acoustically driven spintronic devices based on designer van der Waals heterostructures, which may combine antiferromagnetic, semiconducting, metallic and insulating layers to realise diverse outcomes in spin conversion~\cite{Otani2017,Ghiasi2021}. Moreover, it has been proposed that monolayer CrCl$_3$ exhibits true 2D XY-ferromagnetism, allowing study of the Berezinskii–Kosterlitz–Thouless phase transition~\cite{kosterlitz1973}, and predicted to play host to topological spin textures~\cite{Bedoya2021}. Creation and manipulation of such excitations by SAWs is a tantalising prospect, as has been recently achieved in conventional ferromagnetic systems~\cite{Yokouchi2020}.

\section{Methods}

\subsection{Sample fabrication}

First, IDTs (35 nm aluminium) and electrodes (5 nm titanium / 200 nm gold) are deposited onto 128$\degree$ Y-cut LiNbO$_3$ chips. The IDT fingers are 400 nm wide with 1.2 $\upmu$m spacing, giving a SAW wavelength 3.2 $\upmu$m and frequency 1.1 GHz. The distance between IDT1 and IDT2 is approximately 600 $\upmu$m. Next, bulk CrCl$_3$ is exfoliated onto polydimethylsiloxane (PDMS) sheets using sticky tape (Nitto).  Flakes with uniform thickness are transferred onto LiNbO$_3$ between IDT1 and IDT2 using a conventional PDMS dry stamping technique. Bulk CrCl$_3$ crystals are obtained from the commercial suppliers 2D Semiconductors (USA) and HQ Graphene (Netherlands).

\subsection{Acoustic antiferromagnetic resonance measurements}

The LiNbO$_3$ chip is mounted on a radio-frequency compatible chip carrier and loaded into either a Montana closed-cycle cryostat with external electromagnet in 1 axis (Sample 1), or a helium bath cryostat with superconducting magnet coils in 2 axes (Sample 2). The former has a base temperature around 5~K and the latter around 4.2~K. Both cryostats allow variable sample temperature up to at least 30~K. Coaxial cables are used to connect the chip carrier to a vector network analyzer which is capable of measuring SAW transmission at 1.1 GHz. A time gating function is applied to the signal in order to filter out electromagnetic noise and retrieve the signals S21 and S12 at longer timescales 150 - 250 ns.

\section{Acknowledgements}

The authors would like to thank Joseph Barker, Olena Gomonay, Hidekazu Kurebayashi, and Sean Stansill for helpful comments.
TPL acknowledges support from the JSPS postdoctoral fellowships for research in Japan scheme, and KY from JST PRESTO Grant No.~JPMJPR20LB, Japan and JSPS KAKENHI (No.~21K13886). JP is financially supported by Grants-in-Aid for Scientific Research (S) (No. 19H05629) and JSPS KAKENHI (20H01865), from MEXT, Japan. RSD is supported by Grants-in-Aid for Scientific Research (S) (No. 19H05610), from MEXT, Japan. Y.H. is supported by the RIKEN Junior Research Associate Program. SM is financially supported by JST CREST Grant (No.JPMJCR19J4, No.JPMJCR1874 and No.JPMJCR20C1) and JSPS KAKENH (No.17H02927 and No.20H10865) from MEXT, Japan. YO is is financially supported by Grants-in-Aid for Scientific Research (S) (No. 19H05629).

\section{Author contributions}

T. P. L., J. P. and R. S. D. performed experiments. T. P. L., J. P. and Y. H. fabricated samples. All authors contributed to data interpretation and analysis. K. Y. and S. M. developed the theoretical model. T. P. L. and K. Y. wrote the paper. J. P. and Y. O. initiated and supervised the project.


\bibliography{MagnonPhononBib}







	

\pagebreak
\widetext
\begin{center}
\textbf{\large Supplementary Information for: Acoustically driven magnon-phonon coupling in a layered antiferromagnet}
\end{center}

\maketitle

\section{Supplementary Note 1: Theoretical model}

\subsection{Temperature dependence}

We consider a smiple Heisenberg model of antiferromagnet
\begin{equation}
H = -J_{\parallel } \sum _{\langle m,n\rangle} \left( \bm{S}^A_m \cdot \bm{S}^A_n + \bm{S}_m^B \cdot \bm{S}^B_n \right) - J_{\perp } \sum _{\{ m,n \}} \bm{S}^A_m \cdot \bm{S}_n^B , 
\end{equation}
where $\bm{S}^A_n , \bm{S}^B_n $ denote spins belonging to $A$ and $B$ sublattices respectively with $n$ labelling the lattice sites. For modeling CrCl$_3$, we take $J_{\parallel }>0$ the ferromagnetic intra-layer exchange and $J_{\perp }<0$ the antiferromagnetic inter-layer exchange interactions respectively while $\langle m,n\rangle $ and $\{ m,n\} $ denote intra- and inter-sublattice nearest neighbour links. 

We use the mean-field ansatz
\begin{equation}
\langle \bm{S}_n^A \rangle = \langle S\rangle \hat{\bm{z}}  , \quad \langle \bm{S}_n^B \rangle = -\langle S\rangle \hat{\bm{z}},
\end{equation}
which gives the molecular fields
\begin{equation}
\bm{B}_A = \left( z_{\parallel }J_{\parallel } - z_{\perp }J_{\perp } \right) \langle S\rangle \hat{\bm{z}} , \quad \bm{B}_B = -\left( z_{\parallel }J_{\parallel } -z_{\perp } J_{\perp } \right) \langle S \rangle \hat{\bm{z}} ,
\end{equation}
where  $z_{\parallel }$ and $z_{\perp }$ are the numbers of intra- and inter-sublattice links per atom. The expectation value $\langle S\rangle $ is determined by solving
\begin{equation}
\frac{\langle S\rangle }{S} = B_S \left( q \frac{\langle S\rangle }{S} \right) , \quad q = \frac{z_{\parallel }J_{\parallel } -z_{\perp } J_{\perp }}{k_B T} , \label{eq:mean-field}
\end{equation}
where $B_S \left( x\right) $ is the Brillouin function
\begin{equation}
B_S \left( x\right) = \frac{2S+1}{2S}\coth \frac{2S+1}{2S}x - \frac{1}{2S}\coth \frac{1}{2S}x .
\end{equation}
The asymptotic expansion
\begin{equation}
B_S \left( x\right) = \frac{S+1}{3S}x - \frac{S+1}{3S}\frac{2S^2 +2S+1}{30S^2}x^3 + \cdots , \quad x\rightarrow 0 
\end{equation}
implies at $T=T_N$
\begin{equation}
q_{T_N} =\frac{3S}{S+1} = \frac{z_{\parallel }J_{\parallel } -z_{\perp }J_{\perp }}{k_B T_N}  
\end{equation}
so that one can eliminate the microscopic parameters in favour of $T_N$;
\begin{equation}
q = \frac{3S}{S+1} \frac{T_N}{T} .
\end{equation}
Throughout all the calculations in this work, we take $T_N =14$~K. For the spin parameter $S$, we have two choices; the nominal spin of Cr$^{3+}$ $S=3/2$, or the macrospin approximation $S \rightarrow \infty $. Since past literature report a two-step phase transition where the 2D honeycomb layers first order ferromagnetically, which is followed by the antiferromagnetic order in the out-of-plane direction~\cite{McGuire2017}, we think the latter is more appropriate and use it to compute
\begin{equation}
M_s \left( T\right) = M_s \left( 0\right) \lim _{S\rightarrow \infty } \frac{\langle S\rangle }{S} , \label{eq:Ms}
\end{equation}
where $\langle S\rangle /S$ on the right-hand-side is taken to be the self-consistent (numerical) solution of Eq.~(\ref{eq:mean-field}). We note that the other choice $S=3/2$ gives a similar temperature dependence, and the difference is irrelevant considering the qualitative nature of our analysis. In this formulation, $\bm{B}_A  =-\bm{B}_B $ should be proportional to the exchange field $H_E$ appearing in the spin wave analysis, which yields
\begin{equation}
H_E \left( T \right) = H_E \left( 0\right) \lim _{S\rightarrow \infty } \frac{\langle S\rangle }{S} . \label{eq:HE}
\end{equation}
We fixed the values of $M_s \left( 0\right)$ and $H_E \left( 0\right)$ for Sample 1 such that $M_s \left( T=1.56~{\rm K}\right) = 250$~mT and $H_E \left( T=1.56~{\rm K}\right) = 100$~mT and used them to generate Fig.~1d,e in the main text. Note that the reference temperature of 1.56~K was chosen so as to facilitate comparison with Ref.~\cite{Macneill2019}.

For Sample 2, the uniaxial anisotropy also appears to be important. We model it by adding the following term to the macroscopic free energy density;
\begin{equation}
F_u = - K_u  \left\{  \left( \hat{\bm{u}}\cdot \bm{n}_A \right) ^2 + \left( \hat{\bm{u}}\cdot \bm{n}_B \right) ^2 \right\} ,
\end{equation}
where $\hat{\bm{u}}$ is the unit vector along the easy axis, and $K_u $ is the strength of anisotropy in units of energy density (for definitions of free energy density and $\bm{n}_{A,B}$, see the next section). Defined in this phenomenological way, the temperature dependence of $K_u$ is well established theoretically in a seminal work by Callen~\cite{Callen1966} to be $K_u \propto M_s \left( T\right) ^3$. In computing the transmission spectra, this form was assumed and resulted in the temperature dependence of Fig.~3. 

\subsection{Spin wave resonance fields}

Although the focus of the present work is magneto-elastic coupling, a large part of the experimental results can be understood by considering only magnetic properties of CrCl$_3$. Let $\bm{n}_A = \bm{M}_A /M_s $ and $\bm{n}_B = \bm{M}_B /M_s $ be the normalized magnetization vectors for the respective sublattice, and introduce spherical coordinate variables by
\begin{equation}
\bm{n}_A = \begin{pmatrix}
	\sin \theta _A \cos \phi _A \\
	\sin \theta _A \sin \phi _A \\
	\cos \theta _A \\
	\end{pmatrix} , \quad \bm{n}_B = \begin{pmatrix}
	\sin \theta _B \cos \phi _B \\
	\sin \theta _B \sin \phi _B \\
	\cos \theta _B \\
	\end{pmatrix} .
\end{equation}
We set our coordinate $z$-axis to be along the crystal $c$-axis of CrCl$_3$. All the macroscopic magnetic properties should be derivable from an appropriately constructed free energy density $F$. For our purposes, it is sufficient to assume the following form:
\begin{eqnarray}
F&=& J_E \left\{ \sin \theta _A \sin \theta _B \cos \left( \phi _A -\phi _B \right) + \cos \theta _A \cos \theta _B \right\} - K_{\perp } \left( \cos ^2 \theta _A + \cos ^2 \theta _B \right) \nonumber \\
&& - \frac{K_u}{2} \left\{ \sin ^2 \theta _A \cos 2\left( \phi _A - \phi _u \right) + \sin ^2 \theta _B \cos 2\left( \phi _B -\phi _u \right) \right\}   -\mu _0 M_s \bm{H} \cdot \left( \bm{n}_A + \bm{n}_B \right) .  \label{eq:free_energy}
\end{eqnarray}
Here $J_E >0 $ is the antiferromagnetic exchange energy density (not to be confused with the microscopic exchange energies $J_{\parallel }, J_{\perp }$ in the previous section), $K_{\perp }$ is the out-of-plane uniaxial anisotropy that arises from the intra-layer demagnetizing field and spin-orbit interactions, $K_u \geq 0$ is an externally induced in-plane uniaxial anisotropy that breaks the 6-fold rotation symmetry of CrCl$_3$, $\phi _u $ represents its easy-axis direction that is taken to be a free parameter, and $\bm{H}$ is the external magnetic field. While we assume that the out-of-plane anisotropy is dominated by the demagnetizing field $K_{\perp } = -M_s^2 /2$, the spin-orbit contribution might not be entirely negligible. Although including it can change the theoretical temperature dependence, corrections to the mean-field approximation is far more likely sources of discrepancy so that we do not pursue this direction any further. 

We take $\bm{H}$ to be in the $ab$-plane and parameterize it by
\begin{equation}
\bm{H} = H\begin{pmatrix}
	\cos \phi  \\
	-\sin \phi  \\
	0 \\
	\end{pmatrix} , \quad H>0. \label{eq:angle_convention}
\end{equation}
The unusual sign convention for the $y$-component is in accordance with the clockwise convention of the polar plots in the main text. The equilibrium magnetization configuration is determined by minimizing $F$. It is clear that $\theta _A =\theta _B = \pi /2$. While we speak of spin waves, the wavelength relevant to our study is of order 1 $\mu $m, for which the effect of exchange interactions is expected to be subdominant. Therefore, we treat them as if they were spatially uniform modes. The linearized Landau-Lifshitz equation reads
\begin{equation}
\left\{ i\frac{\omega }{\gamma \mu _0 }\begin{pmatrix}
	0 & -1 & 0 & 0 \\
	1 & 0 & 0 & 0 \\
	0 &0 & 0 & -1 \\
	0 &0 & 1 & 0 \\
	\end{pmatrix} - \begin{pmatrix}
	A_1  & 0 & C_1 & 0 \\
	0 & A_2 & 0 & C_2 \\
	C_1 & 0 & B_1 & 0 \\
	0 & C_2 & 0 & B_2 \\
	\end{pmatrix} \right\} \begin{pmatrix}
	\delta \theta _A \\
	\delta \phi _A \sin \theta _A \\
	\delta \theta _B \\
	\delta \phi _B \sin \theta _B \\
	\end{pmatrix}  =0 ,
\end{equation}
where $\theta _A ,\phi _A ,\theta _B ,\phi _B$ now refer to the equilibrium state and $\delta \theta _A , \delta \phi _A , \delta \theta _B , \delta \phi _B$ small perturbations around it, and
\begin{eqnarray*}
A_1 &=& -H_E \cos \left( \phi _A -\phi _B \right) + H \cos \left( \phi + \phi _A \right) +M_s +H_u \cos 2\left( \phi _A -\phi _u \right) ,\\
A_2 &=& -H_E \cos \left( \phi _A -\phi _B \right) + H \cos \left( \phi + \phi _A \right) +2H_u \cos 2\left( \phi _A -\phi _u \right) , \\
B_1 &=& -H_E \cos \left( \phi _A -\phi _B \right) + H \cos \left( \phi + \phi _B \right) +M_s +H_u \cos 2\left( \phi _B -\phi _u \right) , \\
B_2 &=& -H_E \cos \left( \phi _A -\phi _B \right) + H \cos \left( \phi +\phi _B \right) +2H_u \cos 2\left( \phi _B - \phi _u \right) ,  \\
C_1 &=& H_E , \\
C_2 &=& H_E \cos \left( \phi _A -\phi _B \right) .
\end{eqnarray*}
The eigenfrequencies are obtained to be
\begin{eqnarray}
\frac{\omega ^2}{\gamma ^2 \mu _0^2} &=& \frac{A_1 A_2 + B_1 B_2 + 2C_1 C_2}{2} \pm \Bigg\{ \left( \frac{A_1 A_2 -B_1 B_2}{2} \right) ^2   + \frac{A_1 B_1 + A_2 B_2}{2} \left( C_1^2 +C_2^2 \right) \nonumber \\
&&  + \left( A_1 A_2 + B_1 B_2 \right) C_1 C_2 - \frac{A_1 B_1  - A_2 B_2}{2}  \left( C_1^2 -C_2^2 \right) \Bigg\} ^{1/2} \nonumber \\
&=& \frac{A_1 A_2 + B_1 B_2 +2C_1 C_2}{2} \nonumber \\
&&  \pm \sqrt{\left( \frac{A_1 A_2 -B_1 B_2}{2} \right) ^2 + A_1 B_1 C_2^2 + A_2 B_2 C_1^2 + \left( A_1 A_2 + B_1 B_2 \right) C_1 C_2 }. \label{eq:frequencies}
\end{eqnarray}
Setting $H_u =0$, they reduce to the frequency equivalent of Eq.~(1) in the main text. In generating Fig.~3a in the main text, we minimized $F$ in Eq.~(\ref{eq:free_energy}) numerically to obtain $\phi _A ,\phi _B , \theta _A ,\theta _B$, and evaluated Eq.~(\ref{eq:frequencies}) with the minus sign to plot the acoustic mode resonance frequencies. The parameters for Sample 2, which are temperature dependant via Eqs.~(\ref{eq:Ms}) and (\ref{eq:HE}), were set at $T=1.56$~K as $\mu _0 M_s = 250$~mT, $\mu _0 H_E = 130$~mT, $\mu _0 H_u = 2.4$~mT, and $\phi _u = -\pi /20 $~rad $\approx -9\degree$. Note that $H_E$ being different from Sample 1 is not surprising considering that the two samples were taken from crystals grown in different conditions.

\subsection{Magnon-SAW interactions}

In this subsection, we discuss the coupling between Rayleigh surface acoustic wave and the antiferromagnetic spin waves described in the previous section. Let an isotropic elastic body (the substrate plus the magnetic film on top) occupy the half space $z<0$ and assume there is no stress applied on the surface $z=0$. Acoustic waves in an isotropic media are characterized by only two parameters; longitudinal and transverse sound velocities $c_P$ and $c_S$. When Rayleigh surface acoustic wave with frequency $\omega = c_R k $ is propagating along $x$-axis, the displacement vector $\bm{u}$ is given by
\begin{equation}
\bm{u} = \Re \left[ C  \begin{pmatrix}
	\left( 1-\xi _S^2 \right) \left\{ e^{\kappa _P z} -\left( 1-\xi _S^2 \right) e^{\kappa _S z} \right\}  \\
	0 \\
	-i \sqrt{1-\xi _P^2}\left\{ \left( 1-\xi _S^2 \right) e^{\kappa _P z} -e^{\kappa _S z} \right\} \\
	\end{pmatrix} e^{-i\left( \omega t-kx \right)} \right] .
\end{equation}
Here $c_R $ is the Rayleigh wave velocity solely determined by $c_P ,c_S$, $C$ is a constant, and
\begin{equation}
\kappa _P = k\sqrt{1-\frac{c_R^2}{c_P^2}} , \quad \kappa _S = k\sqrt{1-\frac{c_R^2}{c_S^2}} , \quad \xi _P^2 = \frac{c_R^2}{c_P^2} , \quad \xi _S^2 = \frac{c_R^2}{2c_S^2} .
\end{equation}
The physical discussions should be based on the strain tensor $\epsilon _{ab} = \left( \partial _b u_a + \partial _a u_b \right) /2$ instead of $\bm{u}$ itself. The boundary condition enforces $\epsilon _{zx} =0$ at the surface and it stays close to zero within $\sim 1/k$ from the surface. In our setup, for the $\sim 100$~nm films, we can therefore assume $\epsilon _{zx}$ is subdominant. Then the only nonzero components of the strain tensor to be taken into account are $\epsilon _{xx} $ and $\epsilon _{zz}$. 

We introduce the free energy density of magneto-elasticity to derive magnon-phonon coupling. Assuming full rotational symmetry, we use the following form:
\begin{equation}
F_{\rm me} = b \epsilon _{ab} \left( n_a^A n_b^A  + n_a^B n_b^B \right)  +2c \epsilon _{ab} n_a^A n_b^B . \label{eq:me_free_energy}
\end{equation}
$b$ corresponds to the usual magneto-elastic coupling of ferromagnetic materials while the inter-sublattice coefficient $c$ is peculiar to antiferromagnetic materials. Before going into the detailed calculation, let us see what kind of angular dependence one should expect for the magnon-SAW coupling. First of all, we are interested in the linear dynamics for which the free energy should be quadratic in the small fluctuations. The strain tensor $\epsilon _{ab}$ itself is already a small fluctuation, so that we need to keep only the first order terms in the perturbation of the magnetizations. Denoting the perturbations $\delta n_a^A , \delta n_a^B$ and using $n_a^A , n_b^B$ to refer to the fixed ground state values, the quadratic free energy reads
\begin{equation*}
F_{\rm me} \approx 2b \epsilon _{ab}\left( n_a^A \delta n_b^A + n_a^B \delta n_b^B \right) + 2c \epsilon _{ab} \left( n_a^A \delta n_b^B + n_a^B \delta n_b^A \right) .
\end{equation*}
Next, as far as Rayleigh surface acoustic waves coupled to a thin magnetic field are concerned, as discussed above, we will need to keep only $\epsilon _{xx} $ and $\epsilon _{zz}$. However, because we consider only the ground states in the plane, $n_z^A = n_z^B =0$. Therefore, one can reduce the free energy further to obtain
\begin{equation}
F_{\rm me} \sim 2  \epsilon _{xx} \left\{ b\left( n_x^A \delta n_x^A + n_x^B \delta n_x^B \right) + c \left( n_x^A \delta n_x^B + n_x^B \delta n_x^A \right) \right\} .  \label{eq:me_reduced}
\end{equation}
Let us emphasize that $n_x^A , n_x^B $ are not dynamical variables in this expression but just fixed coefficients that depend on $H$ and $\phi $, which is the root cause of angle dependence of the magnon-phonon couplings. The situation is slightly more complicated, however, since $\bm{n}^A$ and $\delta \bm{n}^A$ (and similarly for $B$) are orthogonal to each other so that some extra angle dependence arises from $\delta n_x^A , \delta n_x^B$. To be quantitative, we write the perturbed magnetisation vectors as
\begin{eqnarray*}
\bm{n}^A &=& \begin{pmatrix}
	\cos \phi _A \\
	\sin \phi _A \\
	0 \\
	\end{pmatrix} , \quad \delta \bm{n}^A \ =\  \delta \phi _A \begin{pmatrix}
	- \sin \phi _A \\
	\cos \phi _A \\
	0 \\
	\end{pmatrix} -\delta \theta _A  \begin{pmatrix}
	0 \\
	0 \\
	1 \\
	\end{pmatrix}  , \\
\bm{n}^B &=& \begin{pmatrix}
	\cos \phi _B \\
	\sin \phi _B \\
	0 \\
	\end{pmatrix} , \quad \delta \bm{n}^B \ = \  \delta \phi _B \begin{pmatrix}
	-\sin \phi _B \\
	\cos \phi _B \\
	0 \\
	\end{pmatrix} - \delta \theta _B \begin{pmatrix}
	0 \\
	0 \\
	1 \\
	\end{pmatrix} .
\end{eqnarray*}
Note that the ground state angles $\phi _A ,\phi _B $ appear not only in the ground state direction but also multiplying the in-plane fluctuations $\delta \phi _A , \delta \phi _B $. Substituting these into Eq.~(\ref{eq:me_reduced}) yields
\begin{eqnarray}
F_{\rm me} &\sim & - 2\epsilon _{xx}  \Big\{ b \left( \delta \phi_A \cos \phi _A \sin \phi _A  + \delta \phi _B\cos \phi _B \sin \phi _B \right) \nonumber \\
&& +c \left( \delta \phi _A \cos \phi _B \sin \phi _A + \delta \phi _B \cos \phi _A \sin \phi _B \right)  \Big\} . \label{eq:me_reduced2}
\end{eqnarray}
If it were a ferromagnet, the direction of magnetization would roughly follow the magnetic field $\phi _A \approx \phi $ so that this expression explains why the magnon-phonon coupling is proportional to $\sin 2\phi = 2\cos \phi \sin \phi $ and maximised at $\phi = 45 \degree $. Since we are dealing with an antiferromagnet, $\phi _A , \phi _B$ have more complicated relations with $\phi $. In addition, the eigenmodes are acoustic and optical (only approximately if $H_u \neq 0$), i.e. in-phase and out-of-phase precessions of $\delta \bm{n}^A$ and $\delta \bm{n}^B$. Thus let us introduce new variables
\begin{equation}
\delta \phi _{\rm ac} = \delta \phi _A + \delta \phi _B , \quad \delta \phi _{\rm op} = \delta \phi _A - \delta \phi _B .
\end{equation}
In terms of those eigenmode variables, Eq.~(\ref{eq:me_reduced2}) reads
\begin{eqnarray}
F_{\rm me} &\sim & -\epsilon _{xx}  \Big[ b \left\{ \delta \phi _{\rm ac} \sin \left( \phi _A + \phi _B \right) \cos \left( \phi _A -\phi _B \right) + \delta \phi _{\rm op} \sin \left( \phi _A -\phi _B \right) \cos \left( \phi _A + \phi _B \right) \right\} \nonumber \\
&& + c \left\{ \delta \phi _{\rm ac} \sin \left( \phi _A + \phi _B \right) + \delta \phi _{\rm op} \sin \left( \phi _A -\phi _B \right) \right\} \Big] ,
\end{eqnarray}
where we used some trigonometric identities to simplify the result. Therefore, in order to understand the angular dependence of the magnon-phonon coupling in antiferromagnets, one needs to know $\phi _A + \phi _B $ and $\phi _A - \phi _B $ as a function of $\phi $. They are in general complicated. However, if there is no in-plane anisotropy, by symmetry considerations, we expect (note our convention of $\phi $ in Eq.~(\ref{eq:angle_convention}))
\begin{equation}
\phi _A + \phi = - \left( \phi _B +\phi \right) \equiv \frac{\pi }{2} - \phi _{\rm cant} .
\end{equation}
The notation is based on the following observation: In the limit of weak field $H\rightarrow 0$, $\bm{n}^A $ and $\bm{n}^B$ are antiparallel to each other and perpendicular to $\bm{H}$ so that the canting angle $\phi _{\rm cant} =0$. $\phi _{\rm cant}$ should monotonically increase as $H$ increases. With this, one obtains
\begin{equation}
\phi _A + \phi _B = -2\phi , \quad \phi _A -\phi _B = \pi - 2\phi _{\rm cant}\left( H\right) .
\end{equation}
Importantly, $\phi _A -\phi _B$ is independent of $\phi $. Thus, one derives
\begin{eqnarray}
g_{\rm ac} & \propto & - \left\{ b\cos \left( \phi _A -\phi _B \right) + c\right\} \sin \left( \phi _A + \phi _B \right)  \ = \ -\left( b\cos 2\phi _{\rm cant} -c  \right) \sin 2\phi , \\
g_{\rm op} &\propto & - \left\{ b\cos \left( \phi _A + \phi _B \right) +c\right\} \sin \left( \phi _A -\phi _B \right) \ = \ - \left( b\cos 2\phi + c \right) \sin 2\phi _{\rm cant} .
\end{eqnarray}
Therefore, the acoustic mode coupling is proportional to $\sin 2\phi $ while the optical coupling is $\cos 2\phi $ for the intra-sublattice term $\propto b$ and angle independent for the inter-sublattice term $\propto c$.

In order to calculate the SAW transmission amplitude, one needs to be more systematic. Following the approach taken by Refs.~\cite{Verba2019,Yamamoto2022} for magnon-SAW coupling in ferromagnetic materials, one may reduce the equations of motion to the following form:
\begin{eqnarray}
\left\{ i\frac{\omega }{\gamma \mu _0} \begin{pmatrix}
	0 & -1 & 0 & 0 \\
	1 & 0 & 0 & 0 \\
	0 & 0 & 0 & -1 \\
	0 & 0 & 1 & 0 \\
	\end{pmatrix} - \begin{pmatrix}
	A_1 & 0 & C_1 & 0 \\
	0 & A_2 & 0 & C_2  \\
	C_1 & 0 & B_1 & 0 \\
	0 & C_2 & 0 & B_2 \\
	\end{pmatrix} \right\} \begin{pmatrix}
	\delta \theta _A \\
	\delta \phi _A \sin \theta _A \\
	\delta \theta _B \\
	\delta \phi _B \sin \theta _B \\
	\end{pmatrix} = \epsilon _R \begin{pmatrix}
	0 \\
	g_A  \\
	0 \\
	g_A \\
	\end{pmatrix} , \label{eq:coupled_1} \\
\left\{ \rho \left( \frac{\omega ^2}{k^2} -c_R^2 \right) + i\eta \omega \right\} \epsilon _R = \sigma + \overline{g_A} \delta \phi _A \sin \theta _A + \overline{g_B} \delta \phi _B \sin \theta _B , \label{eq:coupled_2} 
\end{eqnarray}
where $\epsilon _R$ is an appropriately normalised amplitude of SAW, $\rho = 4650 $ kg/m$^3$ is the mass density of LiNbO$_3$, $c_R \sim 3800 $ m/s is the velocity of Rayleigh mode, $\sigma $ is the external stress generated by the input IDT, and $\eta $ is the coefficient of viscosity in the Kelvin-Voight model of viscoelasticity~\cite{Rose1999}. $g_A$ and $g_B$ are the effective magnon-phonon coupling coefficients in the thin film limit arising from the isotropic magneto-elastic interaction (\ref{eq:me_free_energy}):
\begin{eqnarray}
g_A  &=& -i C_R \sqrt{kd} \left[ b \sin 2\phi _A + c \left\{  \sin \left( \phi _A -\phi _B \right) + \sin \left( \phi _A + \phi _B \right) \right\} \right] , \\
g_B &=& -iC_R \sqrt{kd} \left[ b \sin 2\phi _B + c \left\{ \sin \left( \phi _B -\phi _A \right)  + \sin \left( \phi _A + \phi _B \right) \right\}  \right] , 
\end{eqnarray}
where $C_R$ is a constant of order unity~\cite{Yamamoto2020}. We analytically solved Eqs.~(\ref{eq:coupled_1}) and (\ref{eq:coupled_2}) for $\epsilon _R$ with $\omega = c_R k$, $b=0, C_R \sqrt{kd} c = 10^6$, numerically computed $\phi _A , \phi _B$ for given $H ,\phi $, and evaluated $\epsilon _R$ to generate Figs.~3 and 4. 

As a closing remark, we note that the theoretical model here is meant for capturing qualitative trends. In particular, the theory predicts very sharp lines for optical modes, while the experimental data point to multiple peak structure with a large broadening. There are two main factors that may cause the disagreement:
\begin{enumerate}
    \item The relative height of acoustic and optical peaks depends strongly on the precise form of magneto-elastic free energy, which may contain many more terms than included in Eq.~(\ref{eq:me_free_energy}.
    \item The broadening arising from fluctuations and inhomogeneity is not accounted for, which can become important when the optical mode frequency nears zero.
\end{enumerate}
The growth quality of commercially obtained van der Waals magnetic materials is currently quite poor, but in time the material quality will likely improve, bringing with it our understanding of the above points. For the sake of presentation, the color coding in the simulated polar plots of Figs.~3 and 4 is based on a biased normalization. We use dB units (logarithmic scale), and assign the brightest color to a value of transmission appropriately large compared with the actual minimum of the data set.  This is appropriate given that the figures intend to display the magnetic resonance positions, rather than amplitudes.

\clearpage

\section{Supplementary Note 2: Sample details}

\begin{figure*}[h]
    \center
    \includegraphics[scale=1]{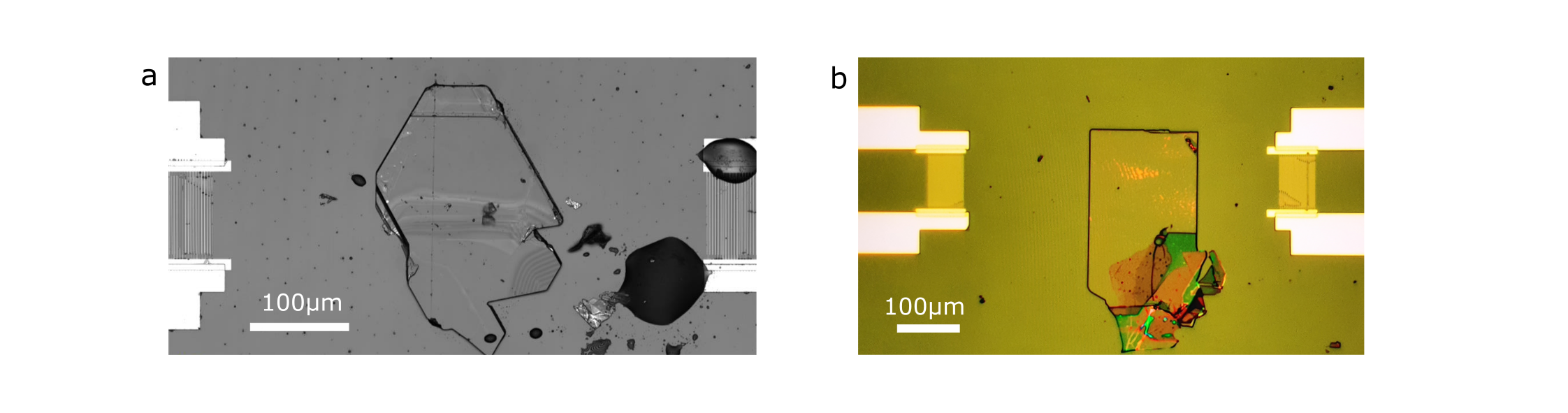}
    \caption{\label{fig:samples} \textbf{Sample images.} (a) Scanning laser microscope image of sample 1, which is $\sim4 \; \upmu$m thick. IDTs can be seen on the left and right of the image. Oil residue on the device was present only after all measurements presented this work were completed. (b) Bright field microscope image of sample 2, which is $\sim 121$ nm thick.}
\end{figure*}

Two samples were studied in this work. Sample 1 is quasi-bulk, at $\sim 4 \; \upmu$m thick (measured by 3D scanning laser microscopy), while sample 2 is much thinner at $\sim 120$ nm (measured by atomic force microscopy). Both flakes were exfoliated with Nitto tape and transferred onto piezoelectric LiNbO$_3$ substrates by polydimethylsiloxane (PDMS) stamping. A laser microscope image of sample 1 is shown in Fig. \ref{fig:samples}a and a bright field microscope image of sample 2 in Fig. \ref{fig:samples}b. 


\section{Supplementary Note 3: Sample 1 response to external magnetic field orientation}

\begin{figure*}[h]
    \center
    \includegraphics[scale=1]{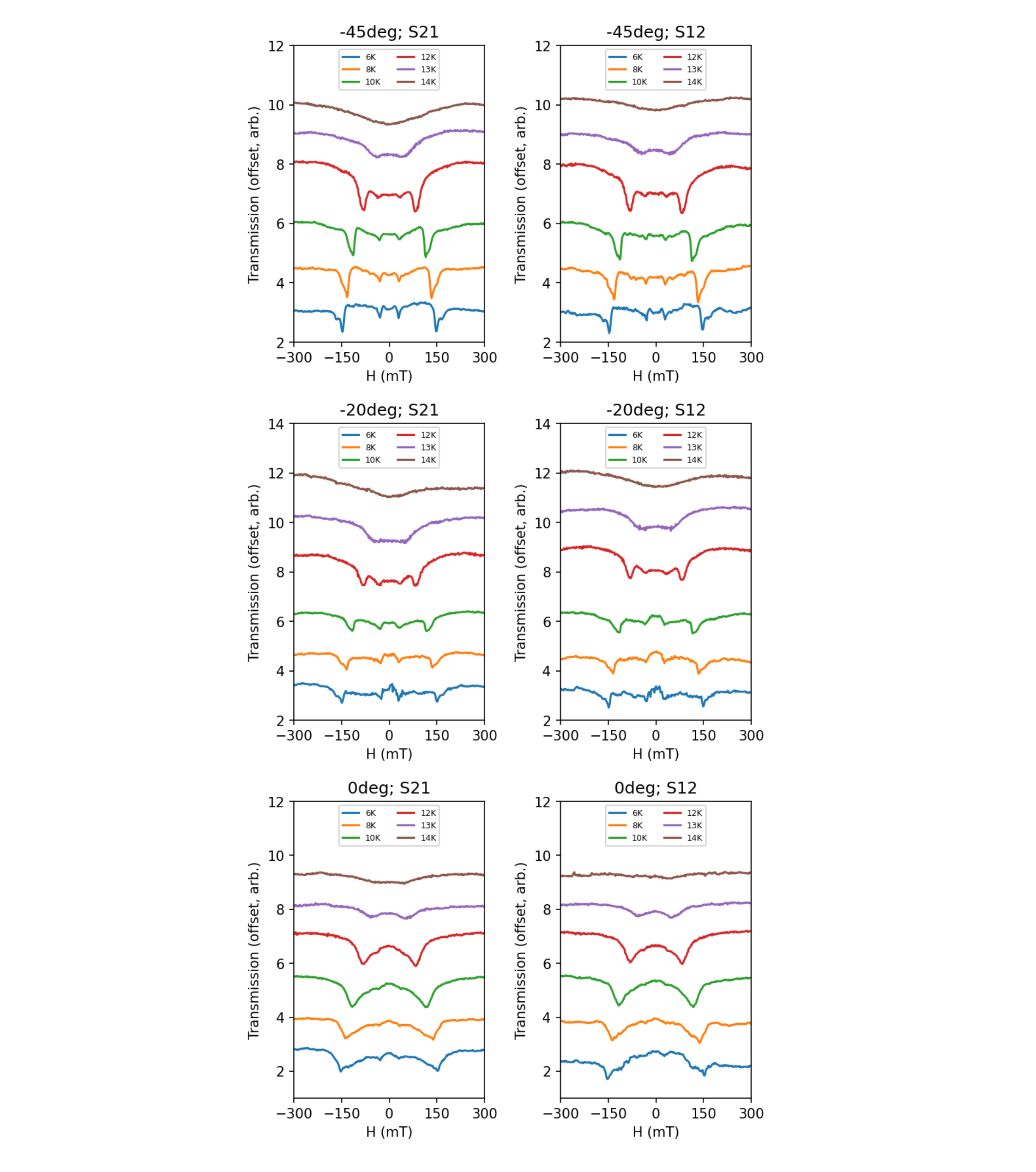}
    \caption{\label{fig:min45_0deg} \textbf{Sample 1 transmission data.} Sample 1 transmission data including S21 and S12 (opposite SAW wavevectors) for the external magnetic field oriented at (a) $-45\degree$ (equivalent to $315\degree$), (b) $-20\degree$ (equivalent to $340\degree$), (c) $0\degree$ to the SAW propagation axis. }
\end{figure*}

\begin{figure*}[h]
    \center
    \includegraphics[scale=1]{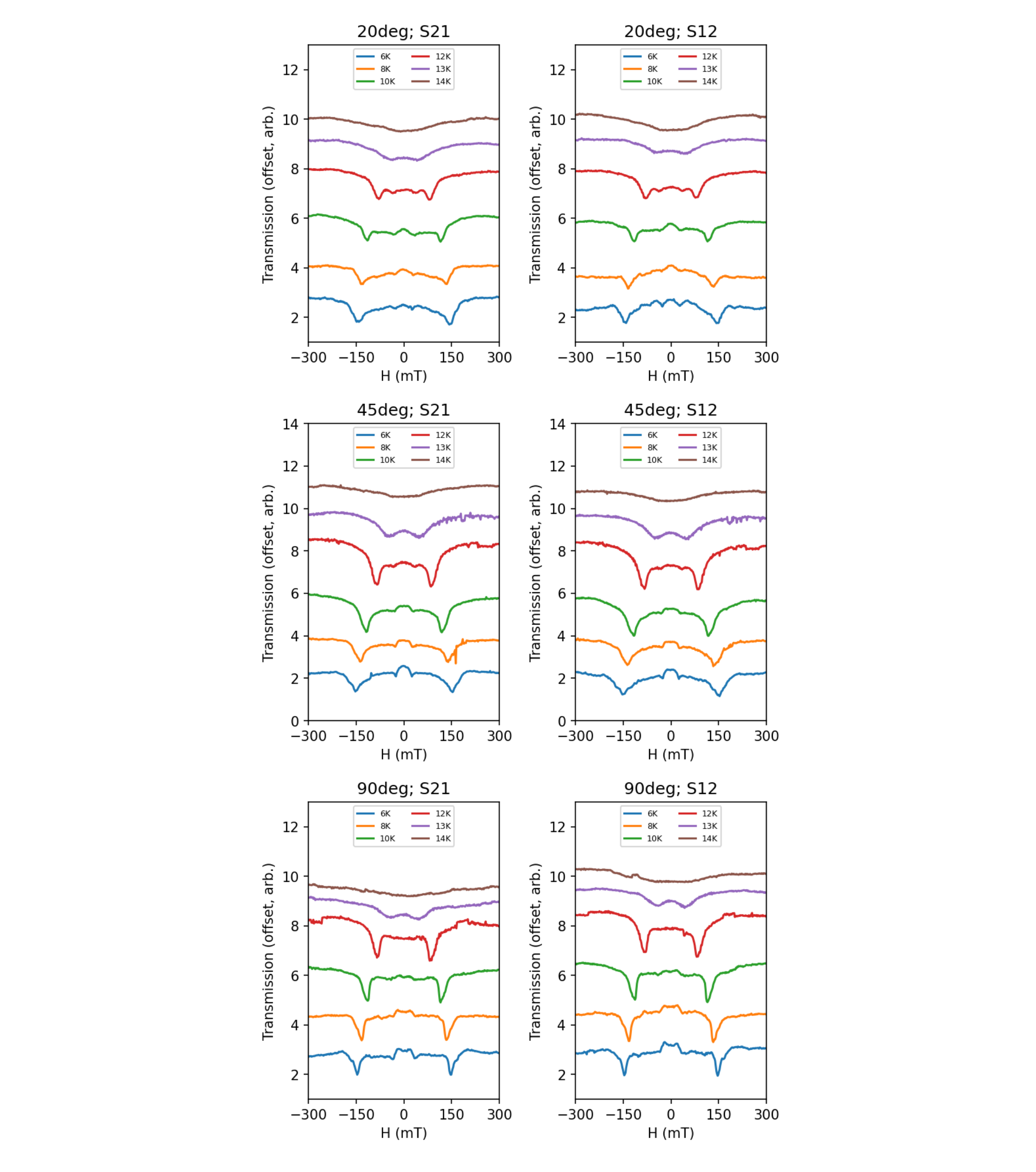}
    \caption{\label{fig:0_90deg} \textbf{Sample 1 transmission data.} Sample 1 transmission data including S21 and S12 (opposite SAW wavevectors) for the external magnetic field oriented at (a) $20\degree$, (b) $45\degree$, (c) $90\degree$ to the SAW propagation axis.}
\end{figure*}

Sample 1 was measured in a Montana magneto-optical cryostat with an electromagnet supplying an external magnetic field in one axis only. Over several repeated cooldown cycles, with manual sample rotation each time, the response of sample 1 to external field orientation can be studied coarsely. Figs \ref{fig:min45_0deg} and \ref{fig:0_90deg} show the SAW transmission in sample 1 at various field orientations. In agreement with sample 2, the optical mode is seen to absorb SAWs at all angles, however, the acoustic mode absorption appears more isotropic in sample 1 compared to sample 2, being present at all angles except for $0\degree$.

\end{document}